\documentclass[prb,aps,twocolumn,superscriptaddress]{revtex4-1}
\usepackage{amsmath}
\usepackage{latexsym}
\usepackage[english]{babel}
\usepackage{graphicx}
\usepackage{bm}
\usepackage{array}
\usepackage{amssymb}
\usepackage{amsfonts}
\usepackage{multirow}
\usepackage{epstopdf}
\usepackage{color}

\begin{document}
	
	\title{Superconductivity driven helical magnetic structure in EuRbFe$_4$As$_4$ ferromagnetic superconductor}
	
	\author{Zh. Devizorova}
	\affiliation{Kotelnikov Institute of Radioengineering and Electronics of RAS, 125009 Moscow, Russia}
	\affiliation{Moscow Institute of Physics and Technology, 141700 Dolgoprudny, Russia}
	\author{A. Buzdin}
	\affiliation{University Bordeaux, LOMA UMR-CNRS 5798, F-33405 Talence Cedex, France}
	\affiliation{Sechenov First Moscow State Medical University, Moscow, 119991, Russia}
	
	\begin{abstract}
		Recently the evidence of the helical magnetic structure modulated along $c$-axis with the period of four lattice parameters was obtained in easy $ab$ plane ferromagnetic superconductor EuRbFe$_4$As$_4$ [K. Iida et al., Phys. Rev. B 100, 014506 (2019)]. We argue that such structure may appear due to the presence of superconductivity. In spite of the very small value of the exchange field acting on the superconducting electrons in EuRbFe$_4$As$_4$, the exchange mechanism of interaction between superconductivity and ferromagnetism could dominate over the electromagnetic one and this circumstance could favor the emergence of the short-period magnetic structure (with the period less than the superconducting coherence length). Such a situation differs from one in the similar compound P-doped EuFe$_2$As$_2$, where the electromagnetic mechanism dominates and results in the magnetic structure with significantly larger period (of the order of London penetration depth). We also analyze the effect of the external magnetic field on the onset temperature of the modulated magnetic structure.
		
	\end{abstract}
	
	\maketitle	
	
	The P-doped EuFe$_2$As$_2$ compound reveals an interesting interplay between superconductivity with $T_c \approx 25$~K and Eu magnetic ordering below $\theta \approx 19$~K \cite{Ahmed_PRL, Cao_JPhys, Nowik_JPhys, Zapf_PRB, Zapf_PRL, Nandi_PRB, Veshchunov_JETP, Stolyarov_SciAdv}. Due to the very low exchange (EX) interaction between Eu moments and conducting electrons, the main mechanism of the superconductivity and magnetism interplay is the electromagnetic (EM) interaction \cite{Devizorova_PRL}. As a result, just below the Curie temperature a domain structure with a period of the order of the London penetration depth emerges instead of ferromagnetic ordering \cite{Krey, Bulaevskii_SSC,Faure_PRL, Dao_PRB, Veshchunov_JETP, Stolyarov_SciAdv, Devizorova_PRL}. With further lowering the temperature it transforms into a usual domain structure while the domains are in the vortex state. Note that the easy axis is along $c$ direction in this compound and the magnetic structure is modulated in the $ab$ plane.
	
	Recently the coexistence of the superconductivity and magnetism was reported in similar EuRbFe$_4$As$_4$ compound with $T_c \approx 37$~K and $\theta \approx 15$~K \cite{Welp_PRB, Stolyarov_PRB}. In contrast with EuFe$_{2}$As$_{2}$, in this case the magnetic moments lie in the $ab$ plane.
	
	Very recent neutron diffraction measurements suggest the helical magnetic structure (HMS) modulated along the $c$-axis in this compound \cite{Iida_PRB}. Moreover, some indications of such HMS were also obtained  by resonant elastic x-ray scattering technique (see Ref.[\onlinecite{Vlasko-Vlasov_PRB_2019}] and reference therein).  Interestingly, the period of this helix appears to be very small, i.e. of the order of four lattice parameters. The recent magnetooptical studies \cite{Vlasko-Vlasov_PRB_2019} indicate the absence of the ferromagnetic vortex phase similar to one observed in EuFe$_2$As$_2$ [\onlinecite{Stolyarov_SciAdv}], which is consistent with a short-period magnetic structure. In Ref. [\onlinecite{Vlasko-Vlasov_PRB_2019}] it was argued that this HMS may have nonsuperconducting origin and caused by a weak antiferromagnetic exchange interaction along the $c$-axis. If this situation is realized, the influence of superconductivity on the helix should be very weak.
	
	Another possibility is that this helix appears due to the presence of superconductivity. At the first glance, it seems that in such the situation the period of the resulting magnetic structure should be basically the same as in similar EuFe$_{2}$As$_{2}$ compound (of the order of the London penetration depth). Since it is significantly larger than four lattice parameters, one can naively expect that the HMS cannot have the superconducting origin. However, there are two important differences between EuFe$_{2}$As$_{2}$ and EuRbFe$_4$As$_4$. First, in these two compounds the magnetic stiffnesses entering the period of the  magnetic structure are essentially different. Indeed, in EuFe$_{2}$As$_{2}$ (EuRbFe$_4$As$_4$)  the magnetic structure is modulated along $c$-axis (in the $ab$-plane) and its period depends on  in-plane (out-of-plane) magnetic stiffness $a_x$ ($a_z$). In these two compounds $a_x$ should be basically the same. At the same time, in EuRbFe$_4$As$_4$ the stiffness $a_z$ is expected to be much smaller than $a_x$ (practically by two order of magnitude). Indeed, since Eu planes are separated by Rb plane, the distance between Eu planes along the $c$-axis is much larger than the distance between Eu atoms in the $ab$ plane. Thus, $a_x$ in EuFe$_{2}$As$_{2}$ could be significantly larger than $a_z$ in EuRbFe$_4$As$_4$. 
	
	Another difference is that the exchange field $h_{EX}$ acting on
	the electrons spin from Eu atoms in EuRbFe$_4$As$_4$ is still small, but substantially larger than that in EuFe$_{2}$As$_{2}$ (see below). 
	
	These circumstances may favor the EX interaction compared  to the EM one in EuRbFe$_4$As$_4$. In such a case, a short-period (with a period less than the superconducting coherence length) magnetic structure \cite{Anderson_PRL} may be realized due to the vanishing of electron spin susceptibility at zero wave-vector in the superconducting state and it could explain the experimental observation of the helix. The spatial profile of this structure should be helical due to easy-plane character of ferromagnetism in EuRbFe$_4$As$_4$. Such structure was first predicted by Anderson and Suhl long time ago in Ref. [\onlinecite{Anderson_PRL}], but never observed before since all previously known ferromagnetic superconductors (FSs) with dominant EX interaction have strong easy-axis magnetic anisotropy \cite{Buzdin_UFN}. The main subject of this Rapid Communication is the detailed analysis of such a situation.

	\begin{figure}[t!]
		\includegraphics[width=0.4\textwidth]{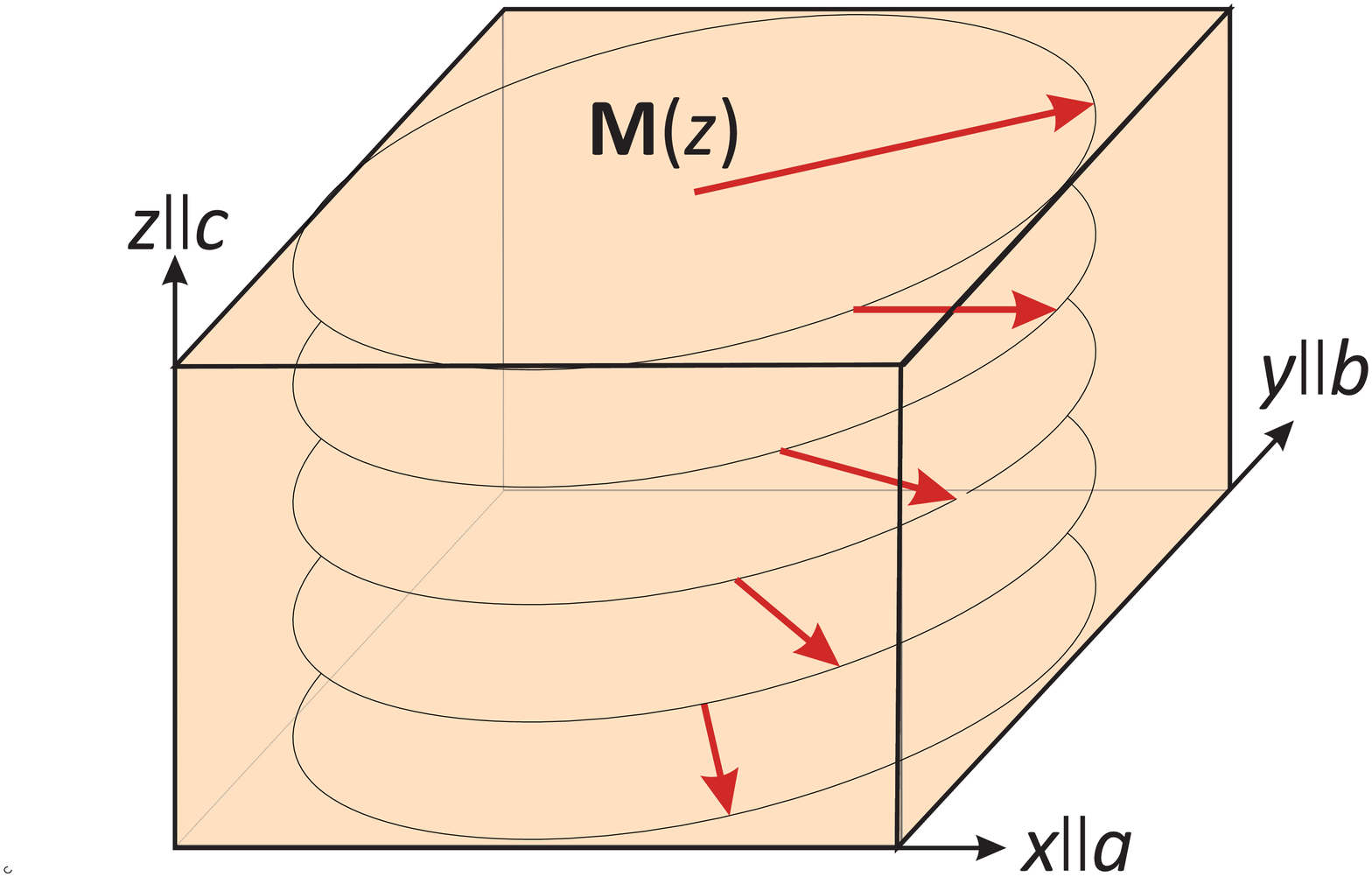}
		\caption{Ferromagnetic superconductor with the magnetization lying in the $ab$ plane and helical magnetic structure modulated along $c$-axis. } \label{sample}
	\end{figure}

	The free energy functional of FS contains the contribution from the superconducting subsystem, the free energy $F_M$ of the magnetic subsystem and the interaction part $F_{int}=F_{int}^{EM}+F_{int}^{EX}$. In the case of easy $xy$-plane ferromagnetism magnetic functional reads as \cite{Buzdin_UFN, Kulic}
	\begin{multline}
	\label{Fm}
	F_M=\frac{({\bf B}-4\pi {\bf M})^2}{8\pi}+
	\frac{n\tilde{\theta}}{M_0^2} \biggl[\tau {\bf M}^2 +a_x^2 \left(\frac{\partial {\bf M}}{\partial x}\right)^2+\\+a_x^2 \left(\frac{\partial {\bf M}}{\partial y}\right)^2 + a_z^2\left(\frac{\partial {\bf M}}{\partial z}\right)^2 + \frac{b{\bf M}^4}{2M_0^2}\biggr]+DM_z^2,
	\end{multline}
	where ${\bf B}$ is the magnetic field, ${\bf M}$ is the magnetization, $n$ is the concentration of magnetic atoms, $\tilde{\theta}$ is the characteristic temperature determined by the mechanism of ferromagnetism, $M_0$ is the saturation magnetization at $T=0$, $\tau=(T-\theta)/\theta$ is the reduced temperature, $\theta$ the Curie temperature, $D$ is out-of-plane anisotropy constant, the coefficient $b \sim 1$.  Since both the EM energy $\theta_{EM}=2\pi M_0^2/n$ and the EX energy $\theta_{EX}=h_{EX}^2 N(E_F)$ are small compared to $\theta$, we assume direct or superexchange interaction to be responsible for the ferromagnetism [here $N(E_F)$ is the electron density of states at the Fermi energy per one localized moment]. In such a case we can put $\tilde{\theta} \approx \theta$. In the absence of the superconductivity ${\bf B}=4\pi {\bf M}$ and the functional Eq.(\ref{Fm}) has the standard form \cite{LL}. The minimum of such functional corresponds to the uniform ferromagnetic state.

	The EM and EX contributions to the interaction energy are the following \cite{Buzdin_UFN, Kulic}
	
	\begin{equation}
	\begin{gathered} 
	F_{int}^{EM}=\sum_{\bf q} K(q)\frac{|{\bf A}_{\bf q}|^2}{2},\\
	F_{int}^{EX}=\sum_{\bf q} \frac{2\pi \theta_{EX}}{\theta_{EM}}\left(\frac{\chi_{n {\bf q}}-\chi_{s {\bf q}}}{\chi_{n 0}}\right)|{\bf M}_{\bf q}|^2 .
	\end{gathered}
	\end{equation}
	Here,  ${\bf A}_{\bf q}$ and ${\bf M}_{\bf q}$ are the Fourier components of the vector-potential and the magnetization, respectively, $K(q)$ is the electromagnetic kernel of a superconductor, which is defined as ${\bf j}_{\bf q}=-c K(q){\bf A}_{\bf q}$, where ${\bf j}_{\bf q}$ is the Fourier component of the Meissner current, $\chi_{n {\bf q}}$ and $\chi_{s {\bf q}}$ are the electronic paramagnetic susceptibilities in the normal and superconducting states, respectively. 
	
	Note that in the presence of superconductivity the magnetic structure becomes nonuniform. Indeed, if we take into account only EX interaction, the minimum of $F_M+F_{int}^{EX}$ will correspond to nonzero wave vector since the paramagnetic susceptibility in the superconducting state $\chi_{s {\bf q}}$ has maximum for $q \ne 0$ [\onlinecite{Anderson_PRL}].
	
	In the clean limit and for $q \xi_0 \ll 1$ the kernel and the susceptibilities read as $K(q)=1/(4\pi \lambda^2)$ and $(\chi_{n {\bf q}}-\chi_{s {\bf q}})/\chi_{n 0}=\left[1-(\pi q \xi_0)^2/18\right]$, while for $q \xi_0 \gg 1$ one has $K(q)=3/(16 \lambda^2 q\xi_0)$  and $(\chi_{n {\bf q}}-\chi_{s {\bf q}})/\chi_{n 0}=\pi/(2q\xi_0)$, where $\xi_0$ is the superconducting coherence length \cite{Kulic}. Below we take $q\xi_0 \gg 1$, but the results remain qualitatively valid for $q \xi_0 \gtrsim 1$.
	
	Since ${\bf B}={\rm rot}{\bf A}$, we solve the Maxwell equation ${\rm rot}({\bf B}-4\pi {\bf M})=4\pi {\bf j}/c$ in the Fourier representation and find  ${\bf A}_{\bf q}=(4\pi i {\bf q} \times {\bf M}_{\bf q})/[q^2+4\pi K(q)]$.
	
	Due to easy {\it ab} plane ferromagnetism in EuRbFe$_4$As$_4$, the magnetic structure in this compound should be helical. We expect this structure to be modulated along $z$-axis since $a_z$ is small compared to $a_x$ (see Fig.\ref{sample}). Choosing the corresponding helical ansatz for the magnetization $M_x=M\cos q_hz$ and $M_y=M\sin q_hz$, we calculate the averaged free energy $\bar{F}=V^{-1}\int F_M dV+F_{int}$:
	\begin{multline}
	\label{FEA_h}
	\bar{F}=2\pi \frac{\theta}{\theta_{EM}}\biggl[\frac{\theta_{EM}}{	 \theta}\frac{1}{1+q_h^2/4\pi K(q_h)} +\tau +a_z^2 q_h^2 +\\+\frac{\theta_{EX}}{\theta} \frac{\pi}{2\xi_0 q_h} \biggr] M^2 +\frac{\theta}{\theta_{EM}}\frac{\pi bM^4}{M_0^2}.
	\end{multline}	
	In the typical case $q_h^2 \gg 4\pi K(q_h)$ we can neglect the unity in denominator of the first term of Eq.(\ref{FEA_h}).
	
	 In the presence of superconductivity, the HMS is stabilized without taking into account interactions such as Dzyaloshinskii-Moriya one. The reason is that in the ferromagnetic state with $q=0$ the first term in (\ref{FEA_h}) reads as $2 \pi(\theta_{EM}+\theta_{EX} +\tau)M^2/\theta_{EM}$ and the increase of $q$ may decrease it. The actual choice of the vector of the helix corresponds to the minimum energy. Similarly, e.g. the HMS in Ho is realized with the modulation vector corresponding to the maximum of the spin susceptibility \cite{Koehler, Chikazumi}.

	Comparing the contributions to the interaction energy coming from EM and EX mechanisms, we find that the later dominates in the coexisting phase formation if $\theta_{EM} < [\theta_{EX} q_h/8 \xi_0 K(q_h)]$. In such a case neglecting the first term in  $\bar{F}$ and minimizing it with respect to $q_h$, one gets the wave vector of the HMS $q_h=(\pi \theta_{EX}/4\theta a_z^2 \xi_0)^{1/3}$, which coincides with the well-known result (see Refs.[\onlinecite{Anderson_PRL, Buzdin_UFN}]). The resulting free energy of the HMS modulated along the $c$-axis reads as:
	
	\begin{equation}
	\label{F_Helix_min}
	\bar{F}_{h}=-\frac{\theta}{\theta_{EM}}\frac{\pi M_0^2}{b}(|\tau|-\tau_h)^2,
	\end{equation}
	where $\tau_h=(\sqrt{27}\pi \theta_{EX} a_z/8\theta \xi_0)^{2/3}$. As a result, the appearance of the helix becomes favorable at $T_h =\theta(1-\tau_h)$.  Note that the magnetic transition temperature is maximal for the modulation along $z$-axis, as $a_z \ll a_x$.
	
	Note that the microscopic model of coexistence between superconductivity and helical magnetism was studied in the pioneering work [\onlinecite{Bulaevskii_LTP}], and the exact solution of the model was obtained.
	
	To confirm the validity of the assumption about the dominance of EX interaction over EM one in EuRbFe$_4$As$_4$, we check the fullfillment of the condition: 
	\begin{equation}
	\label{condition}
	\theta_{EM} < \frac{\theta_{EX} q_h}{8\xi_0 K(q)}.
	\end{equation}

	The exchange field in EuRbFe$_{4}$As$_{4}$ seems to be much larger than in the parent EuFe$_{2}$As$_{2}$ compound [$h_{EX} \sim (0.1-1)$~K \cite{Devizorova_PRL, Pogrebna,Nowik, Nowik2, Jeevan_PRB_08}]. Indeed, the recent measurements \cite{Welp_arxiv} of the upper critical field $H_{c2}$ in EuRbFe$_{4}$As$_{4}$ single crystals show that the orbital critical field $H_{orb}$ at $T=0$~K extrapolated from the $H_{c2}(T)$ slope near $T_{c}$ is around $180$~T for the field orientation in the $ab$ plane, and $100$~T along the $c-$axis. At the same time, the measured \cite{Welp_arxiv} $H_{c2}$ is about $65$~T at low temperatures, which implies the important contribution due to the paramagnetic limitation. The paramagnetic limit  $\mu_{B}H_{P} =1.85\cdot T_{c}\left[K\right]$ in EuRbFe$_{4}$As$_{4}$ is $H_{P} \sim 66$~T, thus the Maki parameter $\alpha_{M}=\sqrt{2}H_{orb}/H_{P}$ is $3.8$ for the
	field orientation in the $ab$ plane and $2.1$ along the $c$-axis. Knowing
	$H_{P}$ and $H_{orb}$ allows us to estimate the theoretical value of $H_{c2}$ at $T=0$~K (see Ref.[\onlinecite{ Gruenberg_PRL_1966}]), which is  $46$~T and $43$~T for $ab$ plane
	and along the $c-$axis, respectively. These values are substantially smaller than the
	experimentally measured $65$~T, which means the weakening of the paramagnetic  effect in EuRbFe$_{4}$As$_{4}$. The possible explanation of such behavior may be related with Jaccarino-Peter effect, i.e. compensation of the field acting on the electrons spins by the exchange field generated by RE atoms \cite{Jaccarino_RPL_1962}. The total field $h=\mu_{B}H\pm\left\vert h_{EX}\right\vert$ may me larger or smaller that the Zeeman field~\ $\mu_{B}H$, depending on the sign of the exchange integral. In some FSs the Jaccarino-Peter effect may lead to the total  compensation of Zeeman field and results in the field-induced  superconductivity. Such situation was observed, for example, in Eu$_{0.75}$Sn$_{0.25}$Mo$_{6}$S$_{7.2}$Se$_{0.8}$ \cite{Meul_RPL_1984}. To explain the observed weakening of the paramagnetic effect in EuRbFe$_{4}$As$_{4}$ we may estimate the value of the exchange field as $h_{EX}\sim (5-10)$~K. Thus, the perturbative treatment of $h_{EX}$ in the $F_{int}$ [see Eq.(\ref{FEA_h})] is justified at all $T < T_h$ due to  $h_{EX} < \Delta$. Taking $h_{EX} \sim 5$ K and $N(E_F) \sim 10$~states/eV per one Eu atom \cite{Stolyarov_PC}, we obtain $\theta _{EX} \sim 2 \times 10^{-2}$~K. 
	
	At the same time, according to recent neutron scattering data \cite{Iida_PRB}, the wave vector of the HMS in EuRbFe$_4$As$_4$ is $q_h \sim 1.2$ nm$^{-1}$. Taking  $\lambda \sim 94$~nm and $\xi_0 \sim 1.2$~nm \cite{Welp_PRB}, we obtain $[\theta_{EX} q_h/8 \xi_0 K(q)] =[2\theta_{EX} q_h^2 \lambda^2/3] \sim 180$~K. On the other hand, we can estimate $\theta_{EM} \sim 1$ K using the parameters from Ref.[\onlinecite{Welp_PRB}]. Thus, the desired condition (\ref{condition}) is fullfield and the EX interaction indeed dominates in EuRbFe$_4$As$_4$.

	Note that  $a_z$ in EuRbFe$_4$As$_4$ is rather small. Indeed, comparing the calculated wave vector $q_h$ with its experimental value \cite{Vlasko-Vlasov_PRB_2019} we obtain $a_z \sim 3 \times 10^{-2}$~nm. Such small value could be realized in EuRbFe$_{4}$As$_{4}$ since the distance between Eu planes along $c$-axis is rather large ($c=13.2$~\AA) and the corresponding exchange integral is small [\onlinecite{Welp_PRB}]. As a result, the small magnetic stiffness and relatively large exchange field in EuRbFe$_{4}$As$_{4}$ favor the dominance of EX interaction. 
	
	In spite the fact that EuRbFe$_4$As$_4$ and EuFe$_{2}$As$_{2}$ compounds seem to be very similar, in the latter one the condition (\ref{condition}) of dominant EX interaction is not fullfield. This violation results from larger magnetic stiffness and smaler exchange energy. Indeed, comparing the period of the modulated magnetic structure in the Meissner state \cite{Devizorova_PRL} with its experimental value \cite{Stolyarov_SciAdv} we obtain $a_x \sim 5$~nm. Taking  $\lambda \sim 350$~nm \cite{Stolyarov_SciAdv}, $\xi_0 \sim 1.5$~nm \cite{Devizorova_PRL, Johnston} and $\theta_{EX} \sim 10^{-3}$~K \cite{Devizorova_PRL}, we can estimate $[\theta_{EX} q_h/8 \xi_0 K(q)]=[\theta_{EX} \pi q_h \lambda^2/2 \xi_0] \sim 1.3$~K. At the same time, the EM energy can be evaluated as $\theta_{EM} \sim 2.1$~K using the parameters from Ref.[\onlinecite{Jeevan_PRB_08}]. Thus, the  condition (\ref{condition}) is violated in EuFe$_{2}$As$_{2}$. Note that the structure of the coexisting phase in this compound is determined by the EM mechanism solely \cite{Devizorova_PRL}.

	In the field cooling regime, the magnetic field lying in the $xy$ plane decreases the temperature of the modulated magnetic structure (MMS) appearance. Indeed, the field ${\bf H}_{ab}=H_{ab} {\bf e}_y$ induces the uniform magnetization ${\bf M}_{ab}=M_{ab} {\bf e}_y$. With lowering the temperature, in addition to this induced magnetization, the MMS ${\bf M}_{m}$ appears. Note that the spatial profile of the structure is not helical now, since the configuration with ${\bf M}_{m} \perp {\bf H}_0$ has lower free energy. As the result, the total magnetization becomes: $M_x=M_m \sin q_hz$, $M_y=M_{ab}$, where $q_h$ is the same as before. Due to the term $\langle M^4\rangle=[M_{ab}^4+(3/8)M_m^4+M_m^2 M_{ab}^2]$ in the free energy, the presence of the induced magnetization renormalizes the onset temperature of the MMS. To calculate the actual temperature shift, below we write down the Gibbs free energy $G=F-{\bf B}{\bf H}_{ab}/4\pi$ corresponding to the above magnetization configuration.
	
	 Note that the strong polarization of the magnetic sybsystem near the Curie temperature leads to the decrease of the London penetration depth and, in principle, may qualitatively change the intervortex interaction \cite{Bespalov_EPL_2015}. EuRbFe$_4$As$_4$ is strong type II superconductor with Ginzburg-Landau parameter  $\kappa \sim 70-100$ [\onlinecite{Welp_PRB}]. Knowing  $H_{orb}$, we may estimate the lower critical field as $H_{c1} \sim 0.2$~kOe. Using the magnetization measurement data at $T \approx 17$~K $\sim \theta$ [\onlinecite{Welp_PRB}], we may estimate the magnetic induction at $H \sim H_{c1}$ as $4\pi M \sim 0.2$~kOe $\sim H_{c1}$. This means practically twofold shrinkage of the effective London penetration depth due to an important contribution to the total field from the magnetization. So we may expect considerable decrease in the vortex size and the increase in the vortex density \cite{Vlasko-Vlasov_PRB_2019} when $T$ approaches $\theta$ from above, which is interesting to observed experimentally. However, this effect practically does not change the value of $H_{c1}$ (see Refs.	[\onlinecite{Bespalov_EPL_2015, Kulic}]) and it is not strong enough to provoke the vortex attraction \cite{Bespalov_EPL_2015}.
	 
	  Neglecting EM contribution to the energy at $q=q_h$, we write down averaged Gibbs free energy at $H_{ab} \gg H_{c1}$ as follows:	
	
	\begin{equation}
	\label{G_m}
	\bar {G}=G_{un}^{ab}+\frac{\pi \theta}{\theta_{EM}} \left(\tau+\tau_h+\frac{b M_{ab}^2}{M_0^2}\right)M_m^2+\frac{3\Gamma b M_m^4}{16M_0^2}.
	\end{equation}
	Here $G_{un}^{ab}$ is the Gibbs energy in the absence of the MMS:
	\begin{multline}
	\label{G_un}
	G_{un}^{ab}=-\frac{H_{ab}^2}{4\pi}+\frac{2\pi \theta}{\theta_{EM}} \biggl( \tau M_{ab}^2+\frac{bM_{ab}^4}{2M_0^2}\biggr)-H_{ab} M_{ab}+\\+2\pi M_{ab}^2+\frac{2\pi \theta_{EX}}{\theta_{EM}}M_{ab}^2,
	\end{multline} 
	where the last term is $F_{int}^{EX}$ at $q=0$. Minimizing Eq.(\ref{G_m}) with respect to $M_m$, we find 
	\begin{equation}
	\label{G_m_min}
	\begin{gathered} 
	M_m^2=\frac{4M_0^2}{3b}\left(|\tau|-\tau_h-\frac{b M_{ab}^2}{M_0^2}\right),\\
	\bar{G}=G_{un}^{ab}-\frac{2\pi \theta}{3\theta_{EM}}\frac{M_0^2}{b}\left(|\tau|-\tau_h-\frac{b M_{ab}^2}{M_0^2}\right)^2.
	\end{gathered} 
	\end{equation}
	
	It follows from Eqs.(\ref{G_m_min}), that in the presence of the magnetization induced by a magnetic field the MMS exists at the temperatures satisfying  $|\tau | \ge [\tau_h +b M_{ab}(\tau)/M_0^2]$. To find the onset temperature $T_m^{ab}=\theta (1-\tau_m^{ab})$ of such structure we solve the equation $\tau_m^{ab} =\tau_h +b M_{ab}(\tau_m^{ab})/M_0^2$ together with the  nonlinear equation for $M_{ab}$:
	\begin{equation}
	\label{Mab}
	\frac{4\pi \theta}{\theta_{EM}}\left[\tau+\tau_{un}+\frac{b M_{ab}^2}{M_0^2}\right]M_{ab}=H_{ab},
	\end{equation}		
	which follows from the condition $\partial G_{un}^{ab}/\partial M_{ab}=0$. Here $\tau_{un}=(\theta_{EM}+\theta_{EX})/\theta$. As a result, we obtain the onset temperature of the MMS in the presence of the external magnetic field:
	
	\begin{equation}
	\label{taumab}
	\tau_m^{ab}=\tau_h+\frac{\theta_{EM}^2b H_{ab}^2}{16\pi^2\theta^2 M_0^2(\tau_{un}-\tau_h)^2}.
	\end{equation}
	$T_m^{ab}$ is always lower than $T_h$.
	
	If the magnetic field exceeds the critical value $H_{cr}$, the formation of the MMS is not favorable. Indeed, it follows from Eq.(\ref{taumab}) that the onset temperature $T_m^{ab}$ of this structure  decreases with increasing the magnetic field. Thus, if the field is above $H_{cr}^{ab}$ [which roughly corresponds to the case $T_m^{ab} \approx 0$ ($\tau_m^{ab} \approx 1$)], the MMS does not appear. The threshold field is: 
	\begin{equation}
	H_{cr}^{ab} \approx\frac{4\pi \theta M_0}{\theta_{EM}\sqrt{b}}(\tau_{un}-\tau_h).
	\end{equation}	
	
	Let us estimate the value $H_{cr}^{ab}$ in EuRbFe$_4$As$_4$. In this compound we have $\tau_h \sim 0.002$, $\tau_{un} \approx \theta_{EM}/\theta \gg  \tau_h$. Thus, $H_{cr}^{ab} \approx 4\pi M_0 /\sqrt{b} \approx 4\pi M_0 \sim 4$~kOe.

	The magnetic field ${\bf H}_c=H_{c}{\bf e}_z$ along $c$-axis also suppreses the onset temperature of the MMS. In this case, the induced magnetization ${\bf M}_c=M_c {\bf e}_z$ is along $z$-axis and obeys the equation:
	
	\begin{equation}
	\label{Mc}
	\frac{4\pi \theta}{\theta_{EM}}\left[\tau+\tau_{un}+\frac{b M_c^2}{M_0^2}+\frac{\theta_{EM}D}{2\pi \theta}\right]M_{c}=H_{c},
	\end{equation}
	which differs from  Eq.(\ref{Mab}) due to the presence of the out-of-plane anisotropy term $D M_c^2$. Since ${\bf M}_c$ does not lie in the $xy$-plane, the MMS should be helical. The total magnetization reads as: $M_x=M_h \cos q_h z$, $M_y=M_h \sin q_h z$, $M_z=M_c$, where $q_h$ is the same as before. Calculating the Gibbs free energy of the above magnetic configuration for $H_c \gg H_{c1}$ we obtain:
	\begin{equation}
	\label{G_m_min_z}
	\bar{G}=G_{un}^c-\frac{\pi \theta}{\theta_{EM}}\frac{M_0^2}{2b}\left(|\tau|-\tau_h-\frac{b M_c^2}{M_0^2}\right)^2.
	\end{equation}
	Thus, the MMS appears at $\tau_m^c =\tau_h +b M_c(\tau_m^c)/M_0^2$. Solving this equation together with the Eq.(\ref{Mc}), we obtain the onset temperature $T_m^c=\theta (1+\tau_m^c)$ of HMS renormalized by an external magnetic field directed along the $c$-axis:
	\begin{equation}
	\label{tau0z}
	\tau_m^c=\tau_h+\frac{\theta_{EM}^2b H_{c}^2}{16 \pi^2\theta^2 M_0^2[\tau_{un}+D\theta_{EM}/(2\pi \theta)-\tau_h]^2}.
	\end{equation}
	The structure does not appear, if the magnetic field exceeds the critical value $H_{cr}^c$:
	\begin{equation}
	H_{cr}^c \approx \frac{4\pi \theta M_0}{\theta_{EM}\sqrt{b}}[\tau_{un}+D\theta_{EM}/(2\pi \theta)-\tau_h].
	\end{equation}	
	
	Let us evaluate the value $H_{cr}^c$ for EuRbFe$_4$As$_4$. To this end, we first estimate the out-of-plane anisotropy constant $D$. According to Ref.[\onlinecite{Vlasko-Vlasov_PRB_2019}] the ratio of out-of-plane $\chi_{\perp}=(\partial M_c/\partial H_c)$ and in-plane $\chi_{\parallel}=(\partial M_{ab}/\partial H_{ab})$ magnetic susceptibilities at $T=(\theta + 1$K) is $\chi_{\perp}/\chi_{\parallel} \sim 1/2$. At the same time, using Eqs.(\ref{Mc}) and (\ref{Mab}), we find this ratio above the Curie temperature:
	
	\begin{equation}
	\frac{\chi_{\perp}}{\chi_{\parallel}}=\frac{\tau +\tau_{un}}{\tau +\tau_{un} +D\theta_{EM}/(2\pi \theta)}.
	\end{equation}
	Since $\theta_{EM} \sim 1$~K, we obtain $D \sim 2$. Thus, we can estimate $H_{cr}^c \sim 5.3$~kOe, which can be easily achieved in the experiments. Note that the indication of the suppression of the MMS onset temperature by the out-of-plane magnetic field was recently observed in Ref.[\onlinecite{Welp_arxiv_2019}].

	To sum up, we have demonstrated that in EuRbFe$_4$As$_4$ compound the small magnetic stiffness along the $c$-axis could make the Ruderman–Kittel–Kasuya–Yosida EX interaction the main mechanism of the interplay between superconductivity and ferromagnetism. This situation differs from one in the similar ferromagnetic superconductor P-doped EuFe$_2$As$_2$, where the EM mechanism dominates. The dominant EX interaction in EuRbFe$_4$As$_4$ may favor the short-period helical magnetic structure with the period $d \propto (a_z^2 \xi_0)^{1/3}$ first predicted by Anderson and Suhl in Ref.[\onlinecite{Anderson_PRL}], but never observed before. The external magnetic field suppresses the onset temperature of the modulated magnetic structure.

	To verify weather the helical structure in EuRbFe$_4$As$_4$ compound is really generated by superconductivity (and not related simply with a maximum of the spin susceptibility at finite wave-vector in the normal state, as e.g. in Ho) it may be interesting to study the effect of pressure. Indeed,  pressure allows one to decrease $T_c$ and increase $T_m$ [\onlinecite{Jackson}]. Thus,  at high pressure we may expect the appearence of the ferromagnetic vortex phase instead of helix, if the predicted scenario is indeed realized.

	\acknowledgements
	
	The authors thank V. Stolyarov and D. Roditchev for fruitful discussions. This work  was supported by the French ANR SUPERTRONICS and OPTOFLUXONICS, EU COST CA16218 Nanocohybri. The work of Zh. D. was carried out within the framework of the state task. Zh. D. is also grateful to the LOMA laboratory for the supporting of her visit to LOMA at University of Bordeaux.

	\end{document}